\documentclass[11pt]{article}

\usepackage{geometry}                					
\usepackage{graphicx, caption, subcaption}
\usepackage{amsmath,amssymb,amsfonts,amsthm,bm}
\usepackage{dsfont}
\usepackage{tikz-cd} 							
\usepackage{enumitem}
\usepackage{cite}
\usepackage{authblk} 							
\usepackage{subfiles} 							
\usepackage{bbm}								
\usepackage{caption}
\usepackage{subcaption}
\usepackage{pifont}

\newcommand{\vol}{\bm{\mathsf{{vol}}}}
\newcommand{\hollowstar}{\text{\ding{73}}}
\newcommand{\msf}[1]{\mathsf{#1}}
\newcommand{\bmsf}[1]{\boldsymbol{\mathsf{#1}}}

\newtheorem{prop}{Proposition}

\title{A mimetic discretization of the macroscopic Maxwell equations in Hamiltonian form}
\author[1]{William Barham}
\author[3]{Yaman G{\"u}{\c c}l{\"u}}
\author[2]{Philip J. Morrison}
\author[3,4]{Eric Sonnendr{\"u}cker}
\affil[1]{Oden Institute for Computational Engineering and Sciences, The University of Texas at Austin}
\affil[2]{Department of Physics and Institute for Fusion Studies, The University of Texas at Austin}
\affil[3]{Max-Plank-Institut f{\"u}r Plasmaphysik, NMPP}
\affil[4]{Technische Universit{\"a}t M{\"u}nchen, Zentrum Mathematik}
\date{\today}                     
\begin{document}

\maketitle
	
\section*{Abstract}
A mimetic spectral element discretization, utilizing a novel Galerkin projection Hodge star operator, of the macroscopic Maxwell equations in Hamiltonian form is presented. The idea of splitting purely topological and metric dependent quantities is natural in the Hamiltonian modeling framework as the Poisson bracket is metric free with the Hamiltonian containing all metric information. This idea may be incorporated into the mimetic spectral element method by directly discretizing the Poincar{\'e} duality structure. This ``split exterior calculus mimetic spectral element method" yields spatially discretized Maxwell's equations which are Hamiltonian and exactly and strongly conserve Gauss's laws. Moreover, the new discrete Hodge star operator is itself of interest as a partition of the purely topological and metric dependent portions of the Hodge star operator. As a simple test case, the numerical results of applying this method to a one-dimensional version of Maxwell's equations are given. 


\section{Introduction} \label{section:intro}
Preservation of the Hamiltonian structure as a priority in structure preserving discretization of PDEs is a relatively recent concept: in plasma physics, structure preserving discretization of the Hamiltonian structure of the Maxwell-Vlasov equations can be found in \cite{Kraus_2017} and \cite{doi:10.1063/1.4962573}; in geophysical fluid dynamics, discretization based on the Hamiltonian structure of the rotating shallow water equations is given in \cite{bauer:hal-02020379}. Each of these methods relies on reformulating the equations of interest in terms of differential forms and exterior calculus. This is done because vector calculus may be elegantly and powerfully reformulated in terms of exterior calculus so that the mathematical origin of physically significant features becomes transparent, and because many powerful methods of discretely representing the structures of exterior calculus have been developed in the past few decades. 

A host of discretization strategies based on exterior calculus (known in different research communities by various aliases: mimetic-, structure-preserving-, and physics-compatible-discretization) have been developed in the past half century. These methods ultimately find their origin in the work of Whitney \cite{whitney_geometric_integration_theory} who defined a map interpolating $k$-cochains to $k$-forms; these interpolated forms have come to be known as Whitney forms. This established a link between the discrete structures of algebraic topology and the continuous world of differential geometry that has proven to be useful in the numerical treatment of PDEs. The relevance of algebraic topology in the context of physical modeling was brought into focus by Tonti in \cite{e_tonti}. The earliest use of Whitney forms in a finite difference method appeared a year later in a paper by Dodziuk \cite{dodziuk_1976}. Early applications of differential forms in computational electrogmanetism can be found in the work of Kotiuga \cite{Kotiuga1985HodgeDA} and Bossavit \cite{bossavit_1988a}, \cite{bossavit_1988b}, and \cite{bossavit_1988c}. Bossavit introduced the use Whitney forms as a basis for finite element discretization and revealed previously unknown connections between mixed finite element methods and algebraic topology. Since this early work, structure preserving discretizations have substantially diversified with representatives including: Mimetic Finite Differences \cite{LipnikovKonstantin2014Mfdm}, Finite Element Exterior Calculus \cite{ArnoldDouglasN2010Feec}, and Discrete Duality Finite Volumes \cite{DDFV}. A discussion of the common features shared among such methods and their often overlooked context in algebraic topology may be found in \cite{bochev_and_hyman}. 

Building on the framework established in \cite{bochev_and_hyman} and the interpolation/histopolation approach of \cite{gerritsma_2010}, \cite{kreeft2011mimetic} created a spectral element discretization based on the double de Rham complex. The distinction of a primal and dual de Rham complex takes into account orientation dependence in a self consistent manner. The explicit treatment of a primal and dual de Rham complex with distinct finite element spaces is likewise found in \cite{bauer:hal-02020379}. Such an approach seeks to separate the purely topological features, e.g. the exterior derivative, from metric dependent features, e.g. the Hodge star operator. The topological features may then be discretized exactly while the metric dependent features incur discretization error. The role of the Hodge star operator in modeling constitutive relations is discussed in \cite{hiptmair_maxwell_cts_disc}, and the construction of discrete Hodge star operators is discussed in \cite{hiptmair_discrete_hodge_star}. 

Structure preserving methods which explicitly discretize the Hodge star operator frequently introduce a dual mesh, see \cite{Desbrun2003DiscreteEC}. Hence, these methods are closely related with staggered grid methods. In fact, highly successful staggered grid methods with celebrated conservation properties such as the Yee scheme \cite{KaneYee1966Nsoi} for electromagnetism or the Arakawa grids \cite{ArakawaAkio1997CDfL} for geophysical fluid dynamics might be understood in terms of structure preserving discretization. See \cite{sonnendrucker_staggered_grid} for a discussion of structure preserving staggered grid methods. As in staggered grid methods, the method developed in this paper identifies certain quantities with a primal grid (straight forms) and others with a staggered grid (twisted forms). 

This paper uses the mimetic spectral element method \cite{kreeft2011mimetic} as a basis to develop a novel Galerkin projection Hodge star operator based on the Poincar{\'e} duality pairing. This allows greater flexibility in separating metric free and metric independent features of physical models. From a modeling perspective, the split exterior calculus framework was used to formulate the continuous theory \cite{eldred_and_bauer}. This approach explicitly separates metric dependent and purely topological quantities in the Hamiltonian formulation of physical theories. Using the mimetic spectral element method augmented with new features based on split exterior calculus, we spatially discretize a Hamiltonian model of the macroscopic Maxwell equations with general (and possibly nonlinear) polarization and magnetization, as introduced in \cite{morrison:GaugeFreeLifting} and formulated in terms of split exterior calculus in \cite{geoMacroMaxwell}. This yields a Hamiltonian system of ODEs which preserve essential features of the continuous system (energy and the Gauss constraints). The numerical results of applying the method to a simple one-dimensional test problem are then presented.

\section{The double de Rham complex and duality structures} \label{section:mimetic_spaces}
The double de Rham complex consists of two de Rham complexes related to each other by the Hodge star operator. The Hodge star operator may be derived from two notions of duality: the $L^2$ inner product and Poincar{\'e} duality. As boundary conditions complicate the notion of Poincar{\'e} duality, only manifolds without boundary are considered. The discretization framework is descended from the landmark work of \cite{bochev_and_hyman}. For our finite element spaces, we employ the mimetic spectral element method \cite{gerritsma_2010} \cite{kreeft2011mimetic}. While discrete Hodge star operators have a long history \cite{bochev_and_hyman} \cite{hiptmair_discrete_hodge_star} \cite{gerritsma_hodge_star}, what distinguishes the method employed here is the explicit use of Poincar{\'e} duality in constructing the discrete Hodge star. The discrete Hodge star operator is the product of two matrices: the inverse $L^2$ mass matrix, and the mass matrix arising from Poincar{\'e} duality. The $L^2$ mass matrix contains all of the metric dependence, while the Poincar{\'e} mass matrix is metric independent. Moreover, the Poincar{\'e} mass matrix yields a discrete integration by parts identity. 

\subsection*{Continuous double de Rham complex}
Let $(\Omega, g)$ be a Riemannian manifold of dimension $n$. Throughout this paper, we let our manifold be the $n$-torus, $\Omega = \mathbb{T}^n$, as all considerations of boundaries are neglected. Let $\{ (V^k, \mathsf{d}_k) \}_{k=0}^n$ be the vector spaces of differential forms on $\Omega$. That is, we define
\begin{equation}
	V^k = H^1 \Lambda^k(\Omega) = \left\{ \eta^k \in L^2 \Lambda^k(\Omega) : \mathsf{d}_k \eta^k \in L^2 \Lambda^{k+1}(\Omega) \right\}. 
\end{equation}
This sequence of vector spaces forms a Hilbert complex \cite{ArnoldDouglasN2010Feec}. For a discussion of the properties of differential forms and definitions of the standard operations on differential forms, e.g. the wedge product and the Hodge star operator, see \cite{kreeft2011mimetic} \cite{ArnoldDouglasN2010Feec}. We may construct a second complex, $\{ (\tilde{V}^k, \tilde{\mathsf{d}}_k) \}_{k=0}^n$, called the complex of twisted forms. This complex is identical to the first in manner of definition, but differs in that twisted forms change sign under orientation changing transformations. These two complexes are related to each other through the Hodge star operator. Diagrammatically, this is given by
\begin{equation}
	\begin{tikzcd}
		\cdots \arrow{r} & V^k \arrow{r}{ \mathsf{d}_k } \arrow{d}{\star} & V^{k+1} \arrow{r} \arrow{d}{\star} & \cdots \\
		\cdots & \tilde{V}^{n-k} \arrow{u}  \arrow{l} & \tilde{V}^{n-(k+1)} \arrow{l}{ \tilde{\mathsf{d}}_{n-(k+1)}} \arrow{u} & \cdots \arrow{l} 
	\end{tikzcd}
\end{equation}
This structure is called the double de Rham complex. In most finite element treatments of exterior calculus, explicit reference to the complex of twisted forms is avoided. Rather, the codifferential operator, the formal $L^2$ adjoint of the exterior derivative, is introduced and those equations which are naturally expressed on the twisted complex are formulated weakly. However, in this work, we elect to explicitly consider both the twisted and straight forms in tandem. 

The Hodge star is typically constructed in a local manner, however for the purposes of this paper it is more convenient to construct the Hodge star as a secondary structure that arises from two different notions of duality on the double de Rham complex. First, we define the $L^2$ inner product of $k$-forms $( \cdot, \cdot ): V^k \times V^k \to \mathbb{R}$ as follows: if $g_x$ is the pointwise inner product on $k$-forms induced by the Riemannian metric and $\vol$ is the volume form, then
\begin{equation}
	( \omega^k, \eta^k) = \int_\Omega g_x( \omega^k, \eta^k) \vol.
\end{equation}
The second notion of duality is Poincar{\'e} duality $\langle \cdot, \cdot \rangle: V^k \times \tilde{V}^{n-k} \to \mathbb{R}$ which is defined
\begin{equation}
	\left\langle \omega^k, \tilde{\eta}^{n-k} \right\rangle = \int_\Omega \omega^k \wedge \tilde{\eta}^{n-k}.
\end{equation}
The $L^2$ inner product is metric dependent while Poincar{\'e} duality is metric independent. This feature will be conserved at the discrete level. So, all quantities in a physical theory which are naturally metric free should be expressed in a manifestly metric free manner with a Poincar{\'e} duality structure and those which are metric dependent with $L^2$ duality. The Hodge star is defined to be the operator $\star: V^k \to \tilde{V}^{n-k}$ such that
\begin{equation}
	\left( \omega^k, \eta^k \right) = \left\langle \omega^k, \star \eta^k \right\rangle.
\end{equation}
The Hodge star is not a single operator, but rather a family of operators. A more precise notation might be $\star_{n-k,k}: V^k \to \tilde{V}^{n-k}$, however we opt for the more concise notation where confusion is unlikely. 

 \subsection*{Finite element double de Rham complex}
As the construction of the finite element spaces on each complex is taken from \cite{kreeft2011mimetic}, we shall be brief in our exposition including only enough detail to establish notation. Suppose that $\mathcal{T}_h$ is a complex of cells defining a discrete representation of the reference element, $[-1,1]^n$. It is defined by dividing $[-1,1]$ into subintervals defined by the nodes $-1 = \xi_0 < \xi_2 < \cdots < \xi_{N} = 1$, and taking tensor products of this $1$-dimensional grid. We let $\mathcal{C}_0$ represent vertices, $\mathcal{C}_1$ edges, $\mathcal{C}_2$ faces, and so on within this complex. We let $| \mathcal{C}_k | = N_k$. We interpret $\bmsf{c} = (\mathsf{c}_i)_{1 \leq i \leq N_k} \in \mathcal{C}_k$ as a vector of numbers which associates numerical coefficients to the geometric entities in the complex. Let $\{ V^k_h \}_{k=0}^n$ be the finite element spaces of differential forms as described in \cite{kreeft2011mimetic}, such that $V^k_h \subset V^k$ and $\text{dim}(V^k_h) = N_k$. 

The decrees of freedom are denoted $\bm{\sigma}^k = (\sigma^k_i)_{1 \leq i \leq N_k}: V^k \to \mathcal{C}_k$. The degrees of freedom is a linear operator which associates a continuous differential $k$-form with coefficients on the chain complex. For $0$-forms, we accomplish this by evaluating pointwise at the vertices. For $1$-forms, we integrate over edges. For $2$-forms, we integrate over faces, and so on. These operators are sometimes called de Rham operators. Moreover, these may be thought of as finite element degrees of freedom.  Interpolation is defined such that $\mathcal{I}^k = \left( \left. \bm{\sigma}^k \right|_{V_h^k} \right)^{-1} : \mathcal{C}_k \to V^k_h$. We denote the basis functions $\{ \Lambda_i^k \}_{i=1}^{N_k}$ so that $\forall \bmsf{c} = (\mathsf{c}_i)_{1 \leq i \leq N_k} \in \mathcal{C}_k$, 
\begin{equation}
	\mathcal{I}^k \bmsf{c} = \sum_{i = 1}^{N_k} \mathsf{c}_i \Lambda^k_i.
\end{equation}
Finally, we define the projection operator $\Pi^k = \mathcal{I}^k \bm{\sigma}^k: V^k \to V^k_h$. Hence,
\begin{equation}
	\Pi^k \phi = \sum_{i=1}^{N_k} \sigma^k_i(\phi) \Lambda^k_i.
\end{equation}
We require $\sigma^k_i ( \Pi^k \phi) = \sigma^k_i (\phi)$ so that the operator is indeed a projection. We assume that
\begin{equation}
	\| \phi - \Pi^k \phi \| = O(h^{p+1})
\end{equation}
where $p \in \mathbb{N}$ is the degree of polynomials used for interpolation. The essential features of the finite element de Rham complex may be summarized in the following commuting diagram:
\begin{equation}
	\begin{tikzcd}
	\cdots \arrow{r} & V^k \arrow{r}{\mathsf{d}_k} \arrow{d}{\bm{\sigma}^k} \arrow[bend right=35,swap]{dd}{\Pi^k} & V^{k+1} \arrow[swap]{d}{\bm{\sigma}^{k+1}} \arrow[bend left=35]{dd}{\Pi^{k+1}} \arrow{r} & \cdots \\
	& \mathcal{C}_k \arrow{r}{\mathbbm{d}_k} \arrow{d}{\mathcal{I}^k} & \mathcal{C}_{k+1} \arrow[swap]{d}{\mathcal{I}^{k+1}} & \\
	\cdots \arrow{r} & V^k_h \arrow{r}{\mathsf{d}_k} & V^{k+1}_h \arrow{r} & \cdots
	\end{tikzcd}
\end{equation}
That the diagram commutes means
\begin{equation}
	\mathbbm{d}_k \bm{\sigma}^k = \bm{\sigma}^{k+1} \mathsf{d}_k, \quad \mathsf{d}_k \mathcal{I}^k = \mathcal{I}^{k+1} \mathbbm{d}_k, \quad \mathsf{d}_k \Pi^k = \Pi^{k+1} \mathsf{d}_k.
\end{equation}
Note, the final expression follows from the first two:
\begin{equation}
	\mathsf{d}_k \Pi^k = \mathsf{d}_k \mathcal{I}^k \bm{\sigma}^k = \mathcal{I}^{k+1} \mathbbm{d}_k \bm{\sigma}^k = \mathcal{I}^{k+1} \bm{\sigma}^{k+1} \mathsf{d}_k = \Pi^{k+1} \mathsf{d}_k.
\end{equation}

We proceed in the same manner beginning from a dual grid, $\tilde{\mathcal{T}}_n$, defined over nodes, $\{ \tilde{\xi}_i \} \subset [-1,1]$, which have been staggered with respect to those of the complex of straight forms. The dual complex is written
\begin{equation}
	\begin{tikzcd}
	\cdots \arrow{r} & \tilde{V}^k \arrow{r}{ \tilde{\mathsf{d}}_k } \arrow{d}{\tilde{\bm{\sigma}}^k} \arrow[bend right=35,swap]{dd}{\tilde{\Pi}^k} & \tilde{V}^{k+1} \arrow[swap]{d}{\tilde{\bm{\sigma}}^{k+1}} \arrow[bend left=35]{dd}{\tilde{\Pi}^{k+1}} \arrow{r} & \cdots \\
	& \tilde{\mathcal{C}}_k \arrow{r}{ \tilde{\mathbbm{d}}_k } \arrow{d}{ \tilde{\mathcal{I}}^k} & \tilde{\mathcal{C}}_{k+1} \arrow[swap]{d}{ \tilde{\mathcal{I}}^{k+1} } & \\
	\cdots \arrow{r} & \tilde{V}^k_h \arrow{r}{ \tilde{\mathsf{d}}_k } & \tilde{V}^{k+1}_h \arrow{r} & \cdots
	\end{tikzcd}
\end{equation}
with all objects defined in a like manner to their analogs in the straight complex. 

 \subsection*{Discrete duality}
Just as the Hodge star couples together the twisted and straight complexes in the continuous setting, so too does a suitably defined discrete Hodge star operator in the discrete context. The construction of this discrete Hodge star operator proceeds analogously to that of the continuous Hodge star operator. The mass matrices associated with the $L^2$-pairing are:
\begin{equation}
	\left( \mathbb{M}_{k} \right)_{ij} = \left( \Lambda^k_i , \Lambda^k_j \right) \quad \text{and} \quad \left( \tilde{\mathbb{M}}_{k} \right)_{ij} = \left( \tilde{\Lambda}^k_i , \tilde{\Lambda}^k_j \right).
\end{equation}

The Poincar{\'e} duality structure is built from the wedge product: $\wedge: \Lambda^k \times \Lambda^l \to \Lambda^{k+l}$ (or $\wedge: \tilde{\Lambda}^k \times \tilde{\Lambda}^l \to \tilde{\Lambda}^{k+l}$). In the discrete setting the wedge product is often associated with the cup product from algebraic topology \cite{WilsonScottO2007Caom}. In general, one should exclusively form the wedge product of forms with like orientation: twisted with twisted and straight with straight. This is because, at the discrete level, twisted forms and straight forms are associated with distinct cell complexes. Hence, mixing these objects haphazardly is ill-advised. However, by considering a Galerkin representation of our discrete twisted and straight forms, we circumvent the usual difficulty coupling the two cell complexes to form a product of twisted and straight forms. That is, we define a family of mass matrices which discretely express Poincar{\'e} duality:
\begin{equation}
	\left( \tilde{\mathbb{M}}_{n-k,k} \right)_{ij} = \left\langle \tilde{\Lambda}^{n-k}_i , \Lambda^k_j \right\rangle \quad \text{and} \quad 
	\left( \mathbb{M}_{n-k,k} \right)_{ij} = \left\langle \Lambda^{n-k}_i , \tilde{\Lambda}^k_j \right\rangle
\end{equation}
where $\{ \Lambda^k_i \}_{i=1}^{N_k}$ are the basis functions for $V^k_h$ and $\{ \tilde{\Lambda}^{n-k}_i \}_{i=1}^{\tilde{N}_{n-k}}$ are the basis functions for $\tilde{V}^{n-k}_h$. We shall frequently call this the Poincar{\'e} mass matrix. 

We may interpret the $L^2$ mass matrix as arising from a dual degrees of freedom operator \cite{Jain2021}. In duality to the primal bases $\{ \Lambda^k_i \}_{i=1}^{N_k} \subset V^k$ and $\{ \tilde{\Lambda}^k_i \}_{i=1}^{\tilde{N}_k} \subset \tilde{V}^k$, we define dual bases of $k$-forms $\{ \Sigma_{i}^k \}_{i=1}^{N_k} \subset V^{k}$ and $\{ \tilde{\Sigma}_i^k \}_{i=1}^{\tilde{N}_k} \subset \tilde{V}^{k}$ such that
\begin{equation}
	\left( \Lambda_i^k, \Sigma_j^k \right) = \delta_{ij} \quad \text{and} \quad \left( \tilde{\Lambda}_i^k, \tilde{\Sigma}_j^k \right) = \delta_{ij}.
\end{equation}
The corresponding dual degrees of freedom, $\bm{\mu}_{i}^k: V^k \to \mathcal{C}_{k}^*$ and $\tilde{\bm{\mu}}_{i}^k: \tilde{V}^k \to \tilde{\mathcal{C}}_{k}^*$, are defined such that 
\begin{equation}
	\bm{\mu}_{i}^k \left( \Sigma_j^k \right) = \delta_{ij} \quad \text{and} \quad \tilde{\bm{\mu}}_{i}^k \left( \tilde{\Sigma}_j^k \right) = \delta_{ij}.
\end{equation}
Hence, it follows that
\begin{equation}
	\bm{\mu}_{i}^k \left( v^k \right) = \left( \Lambda^k_i, v^k \right) \quad \text{and} \quad \tilde{\bm{\mu}}_{i}^k \left( \tilde{u}^k \right) = \left( \tilde{\Lambda}^k_i, \tilde{u}^k \right).
\end{equation}
The dual degrees of freedom, when acting on $v^k_h \in V^k_h \subset V^k$ and $\tilde{u}^k_h \in \tilde{V}^{k}_h \subset \tilde{V}^k$, take the form
\begin{equation}
	\bm{\mu}^k \left( v^k_h \right) = \left( \bm{\Lambda}^k, v^k_h \right) = \mathbb{M}_k \bmsf{v}^k 
	\quad \text{and} \quad 
	\tilde{\bm{\mu}}^k \left( \tilde{u}^k_h \right) = \left( \tilde{\bm{\Lambda}}^k, \tilde{u}^k_h \right) = \tilde{\mathbb{M}}_{k} \tilde{\bmsf{u}}^k.
\end{equation}
Hence, we may interpret $\mathbb{M}_{k}: \mathcal{C}_k \to \mathcal{C}_k^*$ and $\tilde{\mathbb{M}}_{k}: \tilde{\mathcal{C}}_k \to \tilde{\mathcal{C}}_k^*$. 

Likewise, we may interpret the Poincar{\'e} mass matrix as arising from a dual degrees of freedom operator. We define dual bases of $(n-k)$-forms $\{ \tilde{\Sigma}_i^{n-k,k} \}_{i=1}^{N_{n-k}} \subset \tilde{V}^{k}$ and $\{ \Sigma_i^{n-k,k} \}_{i=1}^{\tilde{N}_{n-k}} \subset V^{k}$ such that
\begin{equation}
	\left\langle \Lambda_i^{n-k}, \tilde{\Sigma}_j^{n-k,k} \right\rangle = \delta_{ij} \quad \text{and} \quad \left\langle \tilde{\Lambda}_i^{n-k}, \Sigma_j^{n-k,k} \right\rangle = \delta_{ij}.
\end{equation}
The corresponding dual degrees of freedom, $\tilde{\bm{\mu}}^{n-k,k}_{i}: \tilde{V}^k \to \mathcal{C}_{n-k}^*$ and $\bm{\mu}^{n-k,k}_{i}: V^k \to \tilde{\mathcal{C}}_{n-k}^*$, are defined such that 
\begin{equation}
	\tilde{\bm{\mu}}^{n-k,k}_{i} \left( \tilde{\Sigma}_j^{n-k,k} \right) = \delta_{ij} \quad \text{and} \quad \bm{\mu}^{n-k,k}_{i} \left( \Sigma_j^{n-k,k} \right) = \delta_{ij}.
\end{equation}
Hence, it follows that
\begin{equation}
	\tilde{\bm{\mu}}^{n-k,k}_{i} \left( \tilde{u}^k \right) = \left\langle \Lambda^{n-k}_i, \tilde{u}^k \right\rangle \quad \text{and} \quad \bm{\mu}^{n-k,k}_{i} \left( v^k \right) = \left\langle \tilde{\Lambda}^{n-k}_i, v^k \right\rangle.
\end{equation}
The dual degrees of freedom, when acting on $\tilde{u}^k_h \in \tilde{V}^{k}_h \subset \tilde{V}^k$ and $v^k_h \in V^k_h \subset V^k$, take the form
\begin{equation}
	\begin{aligned}
		\tilde{\bm{\mu}}^{n-k,k} \left( \tilde{u}^k_h \right) &= \left\langle \bm{\Lambda}^{n-k}, \tilde{u}^k_h \right\rangle = \mathbb{M}_{n-k,k} \tilde{\bmsf{u}}^k \\
		\bm{\mu}^{n-k,k} \left( v^k_h \right) &= \left\langle \tilde{\bm{\Lambda}}^{n-k}, v^k_h \right\rangle = \tilde{\mathbb{M}}_{n-k,k} \bmsf{v}^k.
	\end{aligned}
\end{equation}
Hence, these operators act as a projection from the space of $k$-forms onto the degrees of freedom of the twisted $(n-k)$-forms (and vice versa). Hence, $\mathbb{M}_{n-k,k} : \tilde{\mathcal{C}}_k \to \mathcal{C}_{n-k}^*$ and $ \tilde{\mathbb{M}}_{n-k,k}: \mathcal{C}_k \to \tilde{\mathcal{C}}_{n-k}^*$. Whereas the $L^2$ mass matrices are symmetric positive definite, and hence bijections, the Poincar{\'e} mass matrices are not square since in general $N_k \neq \tilde{N}_{n-k}$. 

The following commuting diagram summarizes the projection from the twisted forms into the straight forms:
\begin{equation}
	\begin{tikzcd}[column sep=4em, row sep=2em]
		\mathcal{C}_k \arrow{d}{ \mathbbm{d}_{k}} \arrow{r}{ \mathbb{M}_k }
			& \mathcal{C}_k^* \arrow[shift left=1.5]{l}{ \mathbb{M}_k^{-1} } 
				& \tilde{\mathcal{C}}_{n-k} \arrow{l}{ \mathbb{M}_{k,n-k} }
		\\
		\mathcal{C}_{k+1}  \arrow{r}{ \mathbb{M}_{k+1} } 
			& \mathcal{C}_{k+1}^* \arrow[swap]{u}{ (-1)^{n-k} \mathbbm{d}_k^T}  \arrow[shift left=1.5]{l}{ \mathbb{M}_{k+1}^{-1}} 
				& \tilde{\mathcal{C}}_{n - (k+1)} \arrow[swap]{u}{ \tilde{\mathbbm{d}}_{n-(k+1)} } \arrow{l}{ \mathbbm{M}_{k+1,n-(k+1)} } 
	\end{tikzcd}
\end{equation}
That this diagram commutes is an expression of the discrete integration by parts formula:
\begin{equation}
	\mathbb{M}_{k,n-k} \tilde{\mathbbm{d}}_{n-(k+1)} = (-1)^{n - k} \mathbbm{d}_k^T \tilde{\mathbbm{M}}_{k+1,n-(k+1)}
\end{equation}
which in turn follows from the integration by parts formula:
\begin{equation}
	\left\langle \tilde{\mathsf{d}}_{n-(k+1)} \tilde{\phi}^{n-(k+1)}, \psi^k \right\rangle = (-1)^{n-k} \left\langle \tilde{\phi}^{n-(k+1)}, \mathsf{d}_k \psi^k \right\rangle.
\end{equation}
A similar diagram describes projection from the straight forms into the twisted forms. 

This motivates our definition of the discrete Hodge star operator. At the coefficient level, we define $\hollowstar_{k,n-k}: \tilde{\mathcal{C}}_{n-k} \to \mathcal{C}_k$ and $\tilde{\hollowstar}_{k,n-k}: \mathcal{C}_{n-k} \to \tilde{\mathcal{C}}_k$ as follows:
\begin{equation}
	\begin{aligned}
		\bmsf{u}^k = \hollowstar_{k,n-k} \tilde{\bmsf{u}}^{n-k} = \mathbb{M}_k^{-1} \mathbb{M}_{k,n-k} \tilde{\bmsf{u}}^{n-k} \\
		\tilde{\bmsf{v}}^k = \tilde{\hollowstar}_{k,n-k} \bmsf{v}^{n-k} = \tilde{\mathbb{M}}_k^{-1} \tilde{\mathbb{M}}_{k,n-k} \bmsf{v}^{n-k}.
	\end{aligned}
\end{equation}
At the level of the finite element spaces, we see that the discrete Hodge star operator is identical to that of the continuous Hodge star operator with the trial and test functions restricted to the appropriate finite element spaces:
\begin{equation}
	\begin{aligned}
		u^k_h = \star_{h} \tilde{u}^{n-k}_h \iff \left\langle \eta^k, \tilde{u}^{n-k}_h \right\rangle &= \left( \eta^k, u^k_h \right) \quad \forall \eta^k \in V^k_h \\
		\tilde{v}^k_h = \tilde{\star}_{h} v^{n-k}_h \iff \left\langle \tilde{\chi}^k, v^{n-k}_h \right\rangle &= \left( \tilde{\chi}^k, \tilde{v}^k_h \right) \quad \forall \tilde{\chi}^k \in \tilde{V}^k_h
	\end{aligned}
\end{equation}
where the notation $\star_h$ refers to one of many different discrete Hodge star operators depending on the context. Hence, we see that the discrete Hodge star operator is simply a Galerkin projection. 

\subsection*{Adjoint of degrees of freedom operators}
In the following, we shall show that the reduction and interpolation operators are approximately related through the adjoint operation. 
\begin{prop}
With respect to the Poincar{\'e} duality pairing, $\left( \left. \bm{\sigma}_k \right|_{V^k} \right)^* = \tilde{\mathcal{I}}_{n-k}$. Moreover, 
\begin{equation}
	\frac{ \left| \left\langle ( \bm{\sigma}_k^* - \tilde{\mathcal{I}}_{n-k}) \tilde{\bmsf{u}}, \psi \right\rangle \right|}{ \| \tilde{\bmsf{u}} \| \| \psi \|} \leq \| I - \Pi_k \| \| \tilde{\mathcal{I}}_{n-k} \| \quad \forall \tilde{\bmsf{u}} \in \tilde{\mathcal{C}}^{n-k} \text{ and } \forall \psi \in V^k.
\end{equation} 
Hence, $\tilde{\mathcal{I}}_{n-k}$ approximates $\bm{\sigma}_k^*$ in the following sense:
\begin{equation} \label{adjoint_approx}
	\| \bm{\sigma}_k^* - \tilde{\mathcal{I}}_{n-k} \| := \sup_{ \| \tilde{\bmsf{u}} \| \leq 1 } \sup_{ \| \psi \| \leq 1} \frac{ \left| \left\langle ( \bm{\sigma}_k^* - \tilde{\mathcal{I}}_{n-k}) \tilde{\bmsf{u}}, \psi \right\rangle \right|}{ \| \tilde{\phi}^h \| \| \psi \|} = O(h^{p+1}).
\end{equation}
Similarly, $\mathcal{I}_{n-k}$ approximates $\tilde{\bm{\sigma}}_k^*$.
\end{prop}
\noindent \textbf{Proof:} Let $\psi \in V^k_h$. Then
\begin{equation*}
	\tilde{\bmsf{u}}^T \tilde{\mathbb{M}}_{n-k,k} \bm{\sigma}_k (\psi) = \left\langle \tilde{\mathcal{I}}_{n-k} \tilde{\bmsf{u}}, \mathcal{I}_k \bm{\sigma}_k \psi \right\rangle =  \left\langle \tilde{\mathcal{I}}_{n-k}  \tilde{\bmsf{u}}, \psi \right\rangle
\end{equation*}
since $\mathcal{I}_k \bm{\sigma}_k = \Pi_k = I$ on $V^k_h$. Hence, $\left( \left. \bm{\sigma}_k \right|_{V^k} \right)^* = \tilde{\mathcal{I}}_{n-k}$.

Now, let $\psi \in V^k$ be arbitrary. Then the above tells us
\begin{equation*}
	\tilde{\bmsf{u}}^T \tilde{\mathbb{M}}_{n-k,k} \bm{\sigma}_k (\psi) = \left\langle \tilde{\mathcal{I}}_{n-k} \tilde{\bmsf{u}}, \Pi_k \psi \right\rangle.
\end{equation*}
Hence,
\begin{equation*}
	\left\langle \bm{\sigma}_k^* (\tilde{\bmsf{u}}), \psi \right\rangle - \left\langle \tilde{\mathcal{I}}_{n-k} \tilde{\bmsf{u}}, \psi \right\rangle = - \left\langle \tilde{\mathcal{I}}_{n-k} \tilde{\bmsf{u}}, (I - \Pi_k) \psi \right\rangle 
\end{equation*}
which implies
\begin{equation*}
	\begin{aligned}
		\left| \left\langle \bm{\sigma}_k^* (\tilde{\bmsf{u}}), \psi \right\rangle - \left\langle \tilde{\mathcal{I}}_{n-k} \tilde{\bmsf{u}}, \psi \right\rangle \right|
		&= \left| \left\langle \tilde{\mathcal{I}}_{n-k} \tilde{\bmsf{u}}, (I - \Pi_k) \psi \right\rangle \right| \\
		&\leq \| I - \Pi_k \| \| \tilde{\mathcal{I}}_{n-k} \tilde{\bmsf{u}} \| \| \psi \| \\
		&\leq \| I - \Pi_k \| \| \tilde{\mathcal{I}}_{n-k} \| \| \tilde{\bmsf{u}} \| \| \psi \|.
	\end{aligned}
\end{equation*}
Because $\| I - \Pi_k \| = O(h^{p+1})$ and $\tilde{\mathcal{I}}_{n-k}$, being an operator between finite dimensional spaces, is bounded, (\ref{adjoint_approx}) follows.
 \qed \\
 
A similar result holds for the adjoint with respect to the $L^2$ inner product.
\begin{prop} \label{prop::L2_reduction_adjoint}
With respect to the $L^2$ inner product, $\left( \left. \bm{\sigma}_k \right|_{V^k} \right)^* = \mathcal{I}_k$. Moreover, 
\begin{equation}
	\frac{ \left| \left( ( \bm{\sigma}_k^* - \mathcal{I}_k) \bmsf{u}, \psi \right) \right|}{ \| \bmsf{u} \| \| \psi \|} \leq \| I - \Pi_k \| \| \mathcal{I}_k \| \quad \forall \bmsf{u} \in \mathcal{C}^k \text{ and } \forall \psi \in \Lambda^k.
\end{equation} 
Hence, $\mathcal{I}_k$ approximates $\bm{\sigma}_k^*$ in the following sense:
\begin{equation}
	\| \bm{\sigma}_k^* - \mathcal{I}_k \|:= \sup_{ \| \bmsf{u} \| \leq 1 } \sup_{ \| \psi \| \leq 1} \frac{ \left| \left\langle ( \bm{\sigma}_k^* - \mathcal{I}_k) \bmsf{u}, \psi \right\rangle \right|}{ \| \bmsf{u} \| \| \psi \|} = O(h^{p+1}).
\end{equation}
Similarly, $\tilde{\mathcal{I}}_k$ approximates $\tilde{\bm{\sigma}}_k^*$.
\end{prop}
As we defined $\mathcal{I}^k = \left( \left. \bm{\sigma}^k \right|_{V_h^k} \right)^{-1}$, this implies that the degrees of freedom operator is approximately unitary. 

\subsection*{Relationship between Galerkin projection and Natural Hodge star operators}
The natural Hodge star operator is often used in the mimetic discretization literature \cite{bochev_and_hyman}:
\begin{equation}
	\tilde{\mathbb{H}}_k = \tilde{\bm{\sigma}}_{n-k} \star \mathcal{I}_k \quad \text{and} \quad \mathbb{H}_k = \bm{\sigma}_{n-k} \star \tilde{\mathcal{I}}_k.
\end{equation}

\begin{prop} \label{prop::relating_discrete_hodge}
The Galerkin projection and natural Hodge operators are equal to discretization error
\begin{equation}
	\tilde{\mathbb{H}}_k = \tilde{\hollowstar}_{n-k,k} + O(h^{p+1}).
\end{equation}
Similarly for $\hollowstar_{k,n-k}$ and $\mathbb{H}_{n-k}$. 
\end{prop}
\noindent \textit{Proof:} For all $\tilde{\bmsf{u}} \in \tilde{\mathcal{C}}_{n-k}$ and $\bmsf{v} \in \mathcal{C}_k$, 
\begin{equation}
	\begin{aligned}
		( \tilde{\bmsf{u}} )^T \tilde{\mathbb{M}}_{n-k,n-k} \tilde{\mathbb{H}}_k \bmsf{v}
		&= \tilde{\bmsf{u}}^T \tilde{\mathbb{M}}_{n-k,n-k} \tilde{\bm{\sigma}}_{n-k} \star \mathcal{I}_k \bmsf{v} \\
		&=  ( \tilde{\bm{\sigma}}_{n-k}^* \tilde{\bmsf{u}}, \star \mathcal{I}_k \bmsf{v} ) \\
		&= ( \tilde{\mathcal{I}}_{n-k} \tilde{\bmsf{u}}, \star \mathcal{I}_k \bmsf{v} ) + O(h^p) \\
		&= \left\langle \tilde{\mathcal{I}}_{n-k} \tilde{\bmsf{u}}, (-1)^{k(n-k)} \mathcal{I}_k \bmsf{v} \right\rangle + O(h^p) \\
		&= (-1)^{k(n-k)} (\tilde{\bmsf{u}})^T \tilde{\mathbb{M}}_{n-k,k} \bmsf{v} + O(h^p)
	\end{aligned}
\end{equation}
where we used the result $\| \tilde{\bm{\sigma}}_{n-k}^* - \tilde{\mathcal{I}}_{n-k} \| = O(h^p)$ from proposition \ref{prop::L2_reduction_adjoint} and that $\star \star = (-1)^{k(n-k)}$. Hence,
\begin{equation}
	\tilde{\mathbb{H}}_k = (-1)^{k(n-k)} \tilde{\mathbb{M}}_{n-k,n-k}^{-1} \tilde{\mathbb{M}}_{n-k,k} + O(h^{p+1}).
\end{equation}
\qed

\subsection*{Discrete Functional Derivatives}

Consider a functional $K: \Lambda^k \to \mathbb{R}$. We define two kinds of functional derivatives:
\begin{equation}
	D K[\phi] \delta \phi = \left( \frac{\delta K}{\delta \phi}, \delta \phi \right) = \left\langle \frac{\tilde{\delta} K}{\delta \phi}, \delta \phi \right\rangle
\end{equation}
where $DK[\phi]$ is the standard Fr{\'e}chet derivative. Hence, $\delta K/ \delta \phi \in \Lambda^{k}$ whereas $\tilde{\delta} K/ \delta \phi \in \tilde{\Lambda}^{n-k}$. We call this the latter the twisted functional derivative. See \cite{geoMacroMaxwell} and \cite{eldred:hal-01895935} for more details regarding functional derivatives with respect to this duality pairing. 

\begin{prop} \label{prop::var_deriv_chain_rule}
Let $K: V^k \to \mathbb{R}$. Define $\mathsf{K} := K \circ \mathcal{I}_k : \mathcal{C}_k \to \mathbb{R}$. Then
\begin{equation}
	\frac{\tilde{\delta} K \circ \Pi_k}{\delta u} = \bm{\sigma}_k^* \frac{\tilde{\delta} \mathsf{K}}{\delta \bmsf{u} } = \tilde{\mathcal{I}}_{n-k} \frac{\tilde{\delta} \mathsf{K}}{\delta \bmsf{u}} + O(h^{p+1})
\end{equation}
where $\bmsf{u} = \bm{\sigma}_k u$ and we define
\begin{equation}
	D \mathsf{K}[\bmsf{u}] \delta \bmsf{u} = \left( \frac{\tilde{\delta} \mathsf{K} }{\delta \bmsf{u}} \right)^T \tilde{\mathbb{M}}_{n-k,k} \delta \bmsf{u}.
\end{equation}
If our functional depends only on the finite dimensional space $V^k_h \subset V^k$, i.e. $K: V^k_h \to \mathbb{R}$, then
\begin{equation}
	\frac{\tilde{\delta} K}{\delta \bmsf{u}} = \tilde{\mathcal{I}}_{n-k} \frac{\tilde{\delta} \mathsf{K} }{\delta \bmsf{u} }.
\end{equation}
\end{prop}
\noindent \textbf{Proof:} Notice, $K \circ \Pi_k = \mathsf{K} \circ \bm{\sigma}_k$. Because of this, the result follows from the functional chain rule and prior result for the adjoint of the restriction operator:
\begin{equation*}
	\frac{\tilde{\delta} K \circ \Pi_k}{\delta u} = \bm{\sigma}_k^* \frac{\tilde{\delta} \mathsf{K} }{\delta ( \bm{\sigma}_k(u) ) } = \tilde{\mathcal{I}}_{n-k} \frac{ \tilde{\delta} \mathsf{K} }{\delta \bmsf{u} } + O(h^{p+1}).
\end{equation*}
The equality is exact when $K$ only depends on $V^k_h$ since $\left( \left. \bm{\sigma}_k \right|_{V^k_h} \right)^* = \tilde{\mathcal{I}}_{n-k}$. \qed \\

Suppose that $\bmsf{u}$ only appears in $\mathsf{K}$ as $\tilde{\mathbb{M}}_{n-k,k} \bmsf{u}$. Then
\begin{equation}
	\frac{\partial \mathsf{K}}{\partial \bmsf{u}} = \tilde{\mathbb{M}}_{n-k,k}^T \frac{\partial \mathsf{K}}{\partial \tilde{\mathbb{M}}_{n-k,k} \bmsf{u}}.
\end{equation}
Hence, it follows that in this case,
\begin{equation}
	\frac{\tilde{\delta} \mathsf{K} }{\delta \bmsf{u}} =  \frac{\partial \mathsf{K}}{\partial \tilde{\mathbb{M}}_{n-k,k} \bmsf{u}} =:  \frac{\partial \mathsf{K}}{\partial \bmsf{u}_*}
\end{equation}
where $\bmsf{u}_* = \tilde{\mathbb{M}}_{n-k,k} \bmsf{u} \in \tilde{\mathcal{C}}_{n-k}^*$. Because $\bmsf{u}_* \in \tilde{\mathcal{C}}_{n-k}^*$, it follows that $\partial \msf{K}/ \partial \bmsf{u}_* \in \tilde{\mathcal{C}}_{n-k}$. Note, we could instead alternatively shown that the discrete functional derivative can be obtained as follows:
\begin{equation}
	\frac{\tilde{\delta} ( K \circ \tilde{\mathcal{I}}_{n-k} \circ \tilde{\bm{\mu}}^{n-k,k} ) }{\delta u} =  \tilde{\mathcal{I}}_{n-k} \frac{\partial \mathsf{K}}{\partial \bmsf{u}_*} + O(h^{p+1})
\end{equation}
where $\msf{K} = K \circ \tilde{\mathcal{I}}_{n-k}$ and we interpret $\tilde{\mathcal{I}}_{n-k} \circ \tilde{\bm{\mu}}^{n-k,k}$ as a kind of dual projection. This illuminates the need to define our discrete functional derivatives with respect to the dual variable. Whenever we write derivatives with respect to the variable $\bmsf{u}_*$, we assume that the functional $\msf{K}$ is a functional of $\bmsf{u}_*$ only. 

Similarly, the discrete functional derivative with respect to the $L^2$ duality pairing is given by
\begin{equation} \label{l2_var_deriv}
	\frac{\delta K \circ \Pi_k}{\delta u} = \mathcal{I}_k \frac{\partial \mathsf{K} }{\partial \mathbb{M}_{k,k} \bmsf{u} } + O(h^{p+1}) = \mathcal{I}_k \mathbb{M}_{k,k}^{-1} \frac{\partial \mathsf{K} }{\partial \bmsf{u} } + O(h^{p+1}).
\end{equation}
This is similar to the discrete twisted functional derivative, however we may write this discretized functional derivative with greater flexibility due to the invertibility of the $L^2$ mass matrix. 

To fully discretize the functional derivative, we simply apply the degrees of freedom operator:
\begin{equation}
	\tilde{\bm{\sigma}}_{n-k} \left( \frac{\tilde{\delta} K \circ \Pi_k}{\delta u} \right) = \frac{ \partial \mathsf{K} }{\partial \bmsf{u}_* } + O(h^{p+1})
	\quad \text{and} \quad
	\bm{\sigma}_{k} \left( \frac{\delta K \circ \Pi_k}{\delta u} \right) = \mathbb{M}_k^{-1} \frac{ \partial \mathsf{K} }{\partial \bmsf{u} } + O(h^{p+1})
\end{equation}
since interpolation is the inverse of the degrees of freedom operator.

For the purposes of studying Casimir invariants of the discretized dynamics later, it is helpful to consider variational derivatives of functionals of the form $K[\phi] = \hat{K}[ \mathsf{d} \phi]$. At the continuous level, we have
\begin{equation}
	\left\langle \frac{\tilde{\delta} K}{\delta u}, \delta u \right\rangle = \left\langle \frac{\tilde{\delta} \hat{K}}{\delta (\mathsf{d} u)}, \mathsf{d} \delta u \right\rangle = (-1)^{n - k} \left\langle \mathsf{d} \frac{\tilde{\delta} \hat{K}}{\delta (\mathsf{d} u)}, \delta u \right\rangle + \int_{\partial M} \left( \frac{\tilde{\delta} \hat{K}}{\delta (\mathsf{d} u)} \wedge \delta u \right).
\end{equation}
Hence, under homogeneous boundary conditions or on a manifold without boundary,
\begin{equation}
	\frac{\tilde{\delta} K}{\delta u} = (-1)^{n-k} \mathsf{d} \frac{\tilde{\delta} \hat{K}}{\delta (\mathsf{d} u)}.
\end{equation}
Applying the previous results on discretizing functional derivatives, we find
Let $\mathsf{K} = K \circ \mathcal{I}_k$ and $\hat{\mathsf{K}} = \hat{K} \circ \mathcal{I}_{k+1}$. Then, recalling that $\bmsf{u}_* = \tilde{\mathbb{M}}_{n-k,k} \bmsf{u}$ and $(\mathbbm{d}_k \bmsf{u})_* =  \tilde{\mathbb{M}}_{n-(k+1), k+1} \mathbbm{d}_k \bmsf{u}$, we find
\begin{equation}
	\frac{\partial \mathsf{K}}{\partial \bmsf{u}_*} = (-1)^{n-k} \tilde{\mathbbm{d}}_{n-(k+1)} \frac{\partial \hat{\mathsf{K}} }{\partial (\mathbbm{d}_k \bmsf{u})_* }.
\end{equation}

\section{Discretizing the macroscopic Maxwell equations} \label{section:disc_maxwell}
In \cite{morrison:GaugeFreeLifting} and \cite{geoMacroMaxwell}, a geometrized Hamiltonian theory for the macroscopic Maxwell equations was presented. Here, we briefly review some of those results. 

Suppose $\bm{b}^2 \in V^2$ and $\tilde{\bm{d}}^2 \in \tilde{V}^2$. The Poisson bracket for the macroscopic Maxwell equations is given by
\begin{equation}
	\{F, G\} = 4 \pi c \left[ \left\langle \frac{\tilde{\delta} F}{\delta \tilde{\bm{d}}^2}, \tilde{\mathsf{d}}_1 \frac{\tilde{\delta} G}{\delta \bm{b}^2} \right\rangle - \left\langle \frac{\tilde{\delta} G}{\delta \tilde{\bm{d}}^2}, \tilde{\mathsf{d}}_1 \frac{\tilde{\delta} F}{\delta \bm{b}^2} \right\rangle \right].
\end{equation}
Letting $K: \tilde{V}^2 \times V^2 \to \mathbb{R}$, the Hamiltonian for the macroscopic Maxwell equations is given by
\begin{equation}
	\begin{aligned}
		H[\bm{e}^1, \tilde{\bm{b}}^1] 
			&= K - \int_\Omega \bm{e}^1 \wedge \star \frac{\delta K}{\delta \bm{e}^1} 
				+ \frac{1}{8 \pi} \int_\Omega \left( \bm{e}^1 \wedge \star \bm{e}^1 + \tilde{\bm{b}}^1 \wedge \star \tilde{\bm{b}}^1 \right) \\
			&= K - \left( \bm{e}^1, \frac{\delta K}{\delta \bm{e}^1} \right)
				+ \frac{1}{8 \pi} \left[ \left( \bm{e}^1, \bm{e}^1 \right) + \left( \tilde{\bm{b}}^1, \tilde{\bm{b}}^1 \right) \right]
	\end{aligned}
\end{equation}
where $\tilde{\bm{b}}^1 = \star \bm{b}^2$. We choose to express the Hamiltonian in the variables $(\bm{e}^1, \tilde{\bm{b}}^1)$ because it is necessary to do so in the discretized system. This is a consequence of the non-invertibility of the Poincar{\'e} mass matricies. As the Poisson bracket is a purely topological, metric independent, quantity and the Hamiltonian is metric dependent, we have elected to express each with the appropriate metric independent/dependent structures. This allows our discretized bracket and Hamiltonian to retain these qualities of their continuous counterparts. The constitutive relation is given by
\begin{equation}
	\tilde{\bm{d}}^2 = \tilde{\bm{e}}^2 - 4 \pi \frac{\tilde{\delta} K}{\delta \bm{e}^1} \quad \text{and} \quad \tilde{\bm{h}}^1 = \tilde{\bm{b}}^1 + 4 \pi \frac{\tilde{\delta} K}{\delta \bm{b}^2}.
\end{equation}
We define $\tilde{\bm{e}}^2 = \star \bm{e}^1$, $\tilde{\bm{b}}^1 = \star \bm{b}^2$, $\bm{d}^1 = \star \tilde{\bm{d}}^2$ and $\bm{h}^2 = \star \tilde{\bm{h}}^1$. Moreover, $\bm{e}^1[\tilde{\bm{d}}^2, \bm{b}^2]$ is implicitly defined. 

The gradient of the Hamiltonian is given by
\begin{equation}
	DH[\tilde{\bm{d}}^2, \bm{b}^2](\delta \tilde{\bm{d}}^2, \delta \bm{b}^2) = \frac{1}{4 \pi} \left[ \left\langle \bm{e}^1, \delta \tilde{\bm{d}}^2 \right\rangle + \left\langle \tilde{\bm{h}}^1, \delta \bm{b}^2 \right\rangle \right].
\end{equation}
Hence, the bracket and Hamiltonian for the macroscopic Maxwell equations give rise to the following equations of motion:
\begin{equation}
		\frac{\partial \bm{b}^2}{\partial t} = -c \mathsf{d} \bm{e}^1
		\quad \text{and} \quad
		\frac{\partial \tilde{\bm{d}}^2}{\partial t} = c \tilde{\mathsf{d}} \tilde{\bm{h}}^1.
\end{equation}
The bracket possesses Casimir invariants of the form $F[ \tilde{\mathsf{d}} \tilde{\bm{d}}^2]$ and $F[ \mathsf{d} \bm{b}^2]$ for arbitrary functional $F$.

These are the sourceless Maxwell equations. These may be coupled to a Hamiltonian model for particles, for example a kinetic or fluid model, to obtain a self consistent model for charged particle motion \cite{morrison:GaugeFreeLifting}. Moreover, no boundary conditions are given because we assume periodic boundaries in this paper. 

\subsection*{Discrete macroscopic Maxwell equations}
Directly applying the degrees of freedom operator to Maxwell's equations yields
\begin{equation}
	\begin{split}
		\tilde{\bm{\sigma}}^2 \left( \partial_t \tilde{\bm{d}}^2 \right) &= \tilde{\bm{\sigma}}^2 \left( \mathsf{d} \tilde{\bm{h}}^1 \right) \\
		\bm{\sigma}^2 \left( \partial_t \bm{b}^2 \right) &= \bm{\sigma}^2 \left( - \mathsf{d} \bm{e}^1 \right)
	\end{split}
	\implies
	\begin{split}
		\partial_t \tilde{\bmsf{d}}^2 &= \tilde{\mathbbm{d}}_1 \tilde{\bmsf{h}}^1 \\
		\partial_t \bmsf{b}^2 &= - \mathbbm{d}_1 \bmsf{e}^1.
	\end{split}
\end{equation}
where $\tilde{\bmsf{d}}^2 = \tilde{\bm{\sigma}}^2 \left( \tilde{\bm{d}}^2 \right)$, $\tilde{\bmsf{h}}^1 = \tilde{\bm{\sigma}}^1 \left( \tilde{\bm{h}}^1 \right)$, $\bmsf{b}^2 = \bm{\sigma}^2 \left( \bm{b}^2 \right)$, and $\bmsf{e}^1 = \bm{\sigma}^1 \left( \bm{e}^1 \right)$. In the case of linear media, this is sufficient to have a fully discrete theory. The constitutive relations are merely
\begin{equation}
		\mathbb{M}_{1} \bmsf{e}^1 = \mathbb{M}_{12} \tilde{\bmsf{d}}^2 \quad \text{and} \quad \tilde{\mathbb{M}}_{1} \tilde{\bmsf{h}}^1 = \tilde{\mathbb{M}}_{12} \bmsf{b}^2.
\end{equation}
However, more work is needed to work out the appropriate constitutive relations in nonlinear media. 

As before, we define the dual variables $\bmsf{d}^1_* = \mathbb{M}_{12} \tilde{\bmsf{d}}^2 = \mathbb{M}_1 \bmsf{d}^1 $ and $\tilde{\bmsf{b}}^1_* = \tilde{\mathbb{M}}_{12} \bmsf{b}^2 = \tilde{\mathbb{M}}_1 \tilde{\bmsf{b}}^1$. The Hamiltonian may be written
\begin{equation}
	\begin{aligned}
		H[ \Pi_1 \bm{e}^1, \tilde{\Pi}_1 \tilde{\bm{b}}^1] = \mathsf{H}[\bmsf{e}^1, \tilde{\bmsf{b}}^1] 
			&= \mathsf{K} - \left( \bmsf{e}^1 \right)^T \frac{\partial \mathsf{K}}{\partial \bmsf{e}^1} 
				+ \frac{1}{8 \pi} \left[ (\bmsf{e}^1)^T \mathbb{M}_1 \bmsf{e}^1 + (\tilde{\bmsf{b}}^1)^T \tilde{\mathbb{M}}_1 \tilde{\bmsf{b}}^1 \right] \\
			&= \mathsf{K} - \left( \bmsf{e}^1 \right)^T \frac{\partial \mathsf{K}}{\partial \bmsf{e}^1} 
				+ \frac{1}{8 \pi} \left[ (\bmsf{e}^1)^T \mathbb{M}_1 \bmsf{e}^1 + (\tilde{\bmsf{b}}^1_*)^T \mathbb{M}_1^{-1} \tilde{\bmsf{b}}^1_* \right].
	\end{aligned}
\end{equation}
The discrete bracket is defined as follows:
\begin{equation}
	[ \mathsf{F}, \mathsf{G} ] = \{ F \circ \Pi, G \circ \Pi \}
\end{equation}
where $F \circ \Pi = F[ \tilde{\Pi}_2 \tilde{\bmsf{d}}^2, \Pi_2 \bmsf{b}^2]$. One finds that
\begin{equation}
	\begin{aligned}
		\left[\mathsf{F}, \mathsf{G} \right] 
			&= 4 \pi c \left[ \left\langle \mathcal{I}_1 \frac{\tilde{\delta} \mathsf{F}}{\delta \tilde{\bmsf{d}}^2}, \mathsf{d} \tilde{\mathcal{I}}_1 \frac{\tilde{\delta} \mathsf{G}}{\delta \bmsf{b}^2} \right\rangle 
			- \left\langle \mathcal{I}_1 \frac{\tilde{\delta} \mathsf{G}}{\delta \tilde{\bmsf{d}}^2}, \mathsf{d} \tilde{\mathcal{I}}_1 \frac{\tilde{\delta} \mathsf{F}}{\delta \bmsf{b}^2} \right\rangle \right] \\	
			&= 4 \pi c \left( \frac{\partial \mathsf{F} }{\partial \bmsf{d}^1_* }, \frac{\partial \mathsf{F} }{\partial \tilde{\bmsf{b}}^1_* }\right)^T 
			\begin{pmatrix}
				0 & \mathbb{M}_{12} \tilde{\mathbbm{d}}_1 \\
				- \tilde{\mathbbm{d}}_1^T \tilde{\mathbb{M}}_{21} & 0
			\end{pmatrix}
			\begin{pmatrix}
				\partial \mathsf{G} / \partial \bmsf{d}^1_* \\ 
				\partial \mathsf{G} / \partial \tilde{\bmsf{b}}^1_*
			\end{pmatrix} \\
			&= 4 \pi c \left( \frac{\partial \mathsf{F} }{\partial \bmsf{d}^1_* }, \frac{\partial \mathsf{F} }{\partial \tilde{\bmsf{b}}^1_* }\right)^T 
			\begin{pmatrix}
				0 & \mathbb{M}_{12} \tilde{\mathbbm{d}}_1 \\
				- \tilde{\mathbb{M}}_{12} \mathbbm{d}_1  & 0
			\end{pmatrix}
			\begin{pmatrix}
				\partial \mathsf{G} / \partial \bmsf{d}^1_* \\ 
				\partial \mathsf{G} / \partial \tilde{\bmsf{b}}^1_*
			\end{pmatrix}.
	\end{aligned}
\end{equation}
The discretized constitutive relations are obtained by directly applying the degrees of freedom and using the discrete Hodge star operator to close the system:
\begin{equation}
	\bm{\sigma}^1 ( \bm{d}^1) = \bm{\sigma}^1 (\bm{e}^1) - 4 \pi \bm{\sigma}^1 \left( \frac{\delta K \circ \Pi}{\delta \bm{e}^1} \right) \quad \text{and} \quad \tilde{\bm{\sigma}}_1 ( \tilde{\bm{h}}^1 ) = \tilde{\bm{\sigma}}_1 (\tilde{\bm{b}}^1) + 4 \pi \tilde{\bm{\sigma}}_1 \left( \frac{\tilde{\delta} K \circ \Pi}{\delta \bm{b}^2} \right).
\end{equation}
This simplifies to yield
\begin{equation}
	\bmsf{e}^1 - 4 \pi \mathbb{M}_1^{-1} \frac{\partial \mathsf{K}}{\partial \bmsf{e}^1} = \bmsf{d}^1 
	\quad \text{and} \quad 
	\tilde{\bmsf{h}}^1 = \tilde{\bmsf{b}}^1 + 4 \pi \tilde{\mathbb{M}}_1^{-1} \frac{\partial \mathsf{K}}{\partial \tilde{\bmsf{b}}^1}
\end{equation}
where $\bmsf{d}^1 = \mathbb{M}_1^{-1} \mathbb{M}_{12} \tilde{\bmsf{d}}^2$ and $\tilde{\bmsf{b}}^1 = \tilde{\mathbb{M}}_1^{-1} \tilde{\mathbb{M}}_{12} \bmsf{b}^2$.

\subsection*{Derivatives of the Hamiltonian}
It is easiest to differentiate the Hamiltonian with respect to the variables $(\bmsf{e}^1, \tilde{\bmsf{b}}^1_*)$. Derivatives of $\mathsf{H}[\bmsf{e}^1, \tilde{\bmsf{b}}^1_*]$ are given by:
\begin{equation}
	\frac{\partial \mathsf{H}}{\partial \bmsf{e}^1} = \left( \mathbb{M}_1 - 4 \pi \frac{\partial^2 \mathsf{K}}{\partial \bmsf{e}^1 \partial \bmsf{e}^1} \right) \frac{\bmsf{e}^1}{4 \pi}
	\quad \text{and} \quad 
	\frac{\partial \mathsf{H}}{\partial \tilde{\bmsf{b}}^1_*} = \frac{\partial \mathsf{K}}{\partial \tilde{\bmsf{b}}^1_*} - \left( \frac{\partial^2 \mathsf{K}}{\partial \tilde{\bmsf{b}}^1_* \partial \bmsf{e}^1} \right)^T \bmsf{e}^1 + \frac{ \mathbb{M}_1^{-1} \tilde{\bmsf{b}}^1_*}{4 \pi}.
\end{equation}
Now, we wish to make the transformation $(\bmsf{e}^1, \tilde{\bmsf{b}}^1_*) \mapsto (\bmsf{d}^1_*, \tilde{\bmsf{b}}^1_*)$ since the bracket is in these variables. The Jacobian matrix for this transformation is
\begin{equation}
	\begin{pmatrix}
		\partial \bmsf{d}^1_*/ \partial \bmsf{e}^1 & \partial \bmsf{d}^1_*/ \partial \tilde{\bmsf{b}}^1_*\\
		\partial \tilde{\bmsf{b}}^1_*/ \partial \bmsf{e}^1 & \partial \tilde{\bmsf{b}}^1_*/ \partial \tilde{\bmsf{b}}^1_*
	\end{pmatrix}
	=
	\begin{pmatrix}
		\mathbb{M}_1 - 4 \pi \dfrac{\partial^2 \mathsf{K}}{\partial \bmsf{e}^1 \partial \bmsf{e}^1} & - 4 \pi \dfrac{\partial^2 \mathsf{K}}{\partial \tilde{\bmsf{b}}^1_* \partial \bmsf{e}^1} \\[0.5em]
		0 &\mathbb{I}
	\end{pmatrix}.
\end{equation}
The Jacobian for the inverse transformation $(\bmsf{d}^1_*, \tilde{\bmsf{b}}^1_*) \mapsto (\bmsf{e}^1, \tilde{\bmsf{b}}^1_*)$ is given by
\begin{equation}
	\begin{pmatrix}
		\partial \bmsf{e}^1 / \partial \bmsf{d}^1_* & \partial \tilde{\bmsf{b}}^1_*/ \partial \bmsf{d}^1_* \\
		\partial \bmsf{e}^1 / \partial \tilde{\bmsf{b}}^1_* & \partial \tilde{\bmsf{b}}^1_*/ \partial \tilde{\bmsf{b}}^1_*
	\end{pmatrix}
	=
	\begin{pmatrix}
		\left( \mathbb{M}_1 - 4 \pi \dfrac{\partial^2 \mathsf{K}}{\partial \bmsf{e}^1 \partial \bmsf{e}^1} \right)^{-1} & 4 \pi \left( \mathbb{M}_1 - 4 \pi \dfrac{\partial^2 \mathsf{K}}{\partial \bmsf{e}^1 \partial \bmsf{e}^1} \right)^{-1} \dfrac{\partial^2 \mathsf{K}}{\partial \tilde{\bmsf{b}}^1_* \partial \bmsf{e}^1} \\
		0 & \mathbb{I}
	\end{pmatrix}.
\end{equation}
Hence, noting that
\begin{equation}
	\left( \mathbb{M}_1 - 4 \pi \dfrac{\partial^2 \mathsf{K}}{\partial \bmsf{e}^1 \partial \bmsf{e}^1} \right)^T = \mathbb{M}_1 - 4 \pi \dfrac{\partial^2 \mathsf{K}}{\partial \bmsf{e}^1 \partial \bmsf{e}^1},
\end{equation}
if we let $\bar{\mathsf{H}}[\bmsf{d}^1_*, \tilde{\bmsf{b}}^1_*] = \mathsf{H}[\bmsf{e}^1, \tilde{\bmsf{b}}^1_*]$, then
\begin{equation}
	\frac{\partial \bar{\mathsf{H}}}{ \partial \bmsf{d}^1_*} = \left( \frac{\partial \bmsf{e}^1}{\partial \bmsf{d}^1_*} \right)^T \frac{\partial \mathsf{H}}{\partial \bmsf{e}^1} = \frac{\bmsf{e}^1}{4 \pi}
\end{equation}
and
\begin{equation}
	\frac{\partial \bar{\mathsf{H}}}{ \partial \tilde{\bmsf{b}}^1_*} = \frac{\partial \mathsf{H}}{\partial \tilde{\bmsf{b}}^1_*} + \left( \frac{\partial \bmsf{e}^1}{\partial \tilde{\bmsf{b}}^1_*} \right)^T \frac{\partial \mathsf{H}}{\partial \bmsf{e}^1} = \frac{\tilde{\mathbb{M}}_1^{-1} \tilde{\bmsf{b}}^1_*}{4 \pi} + \frac{\partial \mathsf{K}}{\partial \tilde{\bmsf{b}}^1_*} = \frac{\tilde{\bmsf{h}}^1}{4 \pi}.
\end{equation}
\subsection*{Spatially discretized equations of motion}
The spatially discretized equations of motion may be obtained from the Hamiltonian and bracket using $\dot{F} = \{F, H\}$. Note that both $F$ and $H$ must be expressed as functions of $(\bmsf{d}^1_*, \tilde{\bmsf{b}}^1_*)$. Hence, letting $F = (\mathbb{M}_{12} \tilde{\bmsf{d}}^2, \tilde{\mathbb{M}}_{12} \bmsf{b}^2)$, we find
\begin{equation}
	\begin{aligned}
		\mathbb{M}_{12} \left( \partial_t \tilde{\bmsf{d}}^2 - \tilde{\mathbbm{d}}_1 \tilde{\bmsf{h}}^1 \right) &= 0 \\
		\tilde{\mathbb{M}}_{12} \left( \partial_t \bmsf{b}^2 + \mathbbm{d}_1 \bmsf{e}^1 \right) &= 0.
	\end{aligned}
\end{equation}
If $\mathbb{M}_{12}$ and $\tilde{\mathbb{M}}_{12}$ were invertible, this would be identical to the equations obtained by directly projecting the continuous equations of motion. However, instead, we find that a projected subsystem is Hamiltonian. If the Poincar{\`e} mass matrices are full rank, then we may simply set the inside equal to zero. 

So, we finally find that our discrete system is
\begin{equation}
	\begin{split}
		\partial_t \tilde{\bmsf{d}}^2 &= \tilde{\mathbbm{d}}_1 \tilde{\bmsf{h}}^1 \\
		\partial_t \bmsf{b}^2 &= - \mathbbm{d}_1 \bmsf{e}^1
	\end{split},
	\qquad
	\begin{split}
		&\bmsf{e}^1 - 4 \pi \mathbb{M}_1^{-1} \frac{\partial \mathsf{K}}{\partial \bmsf{e}^1} = \bmsf{d}^1 \\
		&\tilde{\bmsf{h}}^1 = \tilde{\bmsf{b}}^1 + 4 \pi \tilde{\mathbb{M}}_1^{-1} \frac{\partial \mathsf{K}}{\partial \tilde{\bmsf{b}}^1}
	\end{split}
	\qquad \text{and} \qquad
	\begin{split}
		\bmsf{d}^1 &= \mathbb{M}_1^{-1} \mathbb{M}_{12} \tilde{\bmsf{d}}^2 \\
		\tilde{\bmsf{b}}^1 &= \tilde{\mathbb{M}}_1^{-1} \tilde{\mathbb{M}}_{12} \bmsf{b}^2.
	\end{split}
\end{equation}
Moreover, one can immediately see that $\partial_t (\tilde{\mathbbm{d}}_2 \tilde{\bmsf{d}}^2) \equiv 0$ and $\partial_t (\mathbbm{d} \bmsf{b}^2) \equiv 0$ as desired. Hence, the Gauss constraints are conserved. Moreover, both Faraday's and Amp{\`e}re's laws are formulated strongly while the constitutive relations are imposed by Galerkin projection. In finite element treatments of electromagnetism, it is generally the case that one evolution equation is expressed strongly and the other weakly. However, maintaining both the straight and twisted forms in the final expression allows both to be expressed strongly. A similar result is found for Poisson's equation in mixed form in \cite{gerritsma_hodge_star}, although the constitutive relation (the Hodge star operator) is applied differently. 

\subsection*{Temporal discretization}
If $K[\bm{e}^1, \bm{b}^2] = K_e[\bm{e}^1] + K_b[\bm{b}^2]$ (i.e. in cases where the polarization is independent of the magnetic field and magnetization is independent of the electric field), then the standard Hamiltonian splitting procedure for Maxwell's equations works \cite{Kraus_2017}. In more general cases, the approach to temporal discretization must be handled case by case. 

We split the Hamiltonian as follows. Define
\begin{equation}
	H_e = K_e - \left( \bm{e}^1, \frac{\delta K_e}{\delta \bm{e}^1} \right) + \frac{1}{8 \pi} \left( \bm{e}^1, \bm{e}^1 \right)
\end{equation}
and
\begin{equation}
	H_b = K_b + \frac{1}{8 \pi} \left( \bm{b}^2, \bm{b}^2 \right).
\end{equation}
Then one finds that $H = H_e + H_b$ and
\begin{equation}
	\frac{ \tilde{\delta} H_e}{\delta \tilde{\bm{d}}^2 } = \frac{\bm{e}^1}{4 \pi} \quad \text{and} \quad \frac{\tilde{\delta} H_b}{\delta \bm{b}^2} = \frac{\tilde{\bm{h}}^1}{4 \pi}
\end{equation}
while $\tilde{\delta} H_e/ \delta \bm{b}^2 = \tilde{\delta} H_b/ \delta \tilde{\bm{d}}^2 = 0$. Hence, the Hamiltonians $H_e$ and $H_b$ give rise to the equations of motion
\begin{equation}
	\begin{aligned}
		\partial_t \tilde{\bm{d}}^2 &= c \mathsf{d} \tilde{\bm{h}}^1 \\
		\partial_t \bm{b}^2 &= 0
	\end{aligned}
	\quad \text{and} \quad
	\begin{aligned}
		\partial_t \tilde{\bm{d}}^2 &= 0 \\
		\partial_t \bm{b}^2 &= - c \mathsf{d} \bm{e}^1
	\end{aligned}
\end{equation}
respectively. We compose the two flows together using Hamiltonian splitting formulas derived from the Baker-Campbell-Hausdorff formula \cite{geometric_numerical_integration}.

\section{Numerical Experiments in 1D} \label{section:numerical_exp_1d}
For ease of implementation, we restrict ourselves to consider Maxwell's equations in a single spatial dimension and periodic boundary conditions. In this case, Maxwell's equations become
\begin{equation}
	\partial_t b^1 = - c \mathsf{d}_0 e^0 \quad \text{and} \quad \partial_t \tilde{d}^1 = c \tilde{\mathsf{d}}_0 \tilde{h}^0,
\end{equation}
the constitutive relations become
\begin{equation}
	\tilde{d}^1 = \tilde{e}^1 - 4 \pi \frac{\tilde{\delta} K}{\delta e^1} \quad \text{and} \quad \tilde{h}^0 = \tilde{b}^0 + 4 \pi \frac{\tilde{\delta} K}{\delta b^1},
\end{equation}
the Poisson bracket becomes
\begin{equation}
	\{F, G\} = 4 \pi c \left[ \left\langle \frac{\tilde{\delta} F}{\delta \tilde{d}^1}, \tilde{\mathsf{d}}_0 \frac{\tilde{\delta} G}{\delta b^1} \right\rangle - \left\langle \frac{\tilde{\delta} G}{\delta \tilde{d}^1}, \tilde{\mathsf{d}}_0 \frac{\tilde{\delta} F}{\delta b^1} \right\rangle \right],
\end{equation}
and the Hamiltonian becomes
\begin{equation}
	H = K - \left( e^0, \frac{\delta K}{\delta e^0} \right) + \frac{1}{8 \pi} \left[ \left( e^0, e^0 \right) + \left( b^1, b^1 \right) \right].
\end{equation}

\subsection*{Discretized equations}
Proceeding as in the 3D case, we find the discrete equations
\begin{equation}
	\begin{split}
		\partial_t \tilde{\msf{d}}^1 &= c \tilde{\mathbbm{d}}_0 \tilde{\msf{h}}^0 \\
		\partial_t \msf{b}^1&= - c \mathbbm{d}_0 \msf{e}^0
	\end{split},
	\qquad
	\begin{split}
		&\msf{e}^0 - 4 \pi \mathbb{M}_0^{-1} \frac{\partial \mathsf{K}}{\partial \msf{e}^0} = \msf{d}^0 \\
		&\tilde{\msf{h}}^0 = \tilde{\msf{b}}^0 + 4 \pi \tilde{\mathbb{M}}_0^{-1} \frac{\partial \mathsf{K}}{\partial \tilde{\msf{b}}^0}
	\end{split}
	\qquad \text{and} \qquad
	\begin{split}
		\msf{d}^0 &= \mathbb{M}_0^{-1} \mathbb{M}_{01} \tilde{\msf{d}}^1 \\
		\tilde{\msf{b}}^0 &= \tilde{\mathbb{M}}_0^{-1} \tilde{\mathbb{M}}_{01} \msf{b}^1.
	\end{split}
\end{equation}
As in the previous case, these discretized equations are a Hamiltonian system with
\begin{equation}
	\msf{H}(\msf{e}^0, \msf{b}^1) = \msf{K} - ( \msf{e}^0 )^T \frac{\partial \msf{K}}{\partial \msf{e}^0} + \frac{1}{8 \pi} \left[ ( \msf{e}^0 )^T \mathbb{M}_0 \msf{e}^0 + ( \msf{b}^1 )^T \mathbb{M}_1 \msf{b}^1 \right].
\end{equation}
In each example, we use linear constitutive relations.

\subsection*{Overview of the numerical method in 1D}
The finite element spaces used are identical to those developed in \cite{kreeft2011mimetic}. Hence, we refrain from including any details of their construction except those that are necessary. The method uses the interpolation/histopolation approach of \cite{gerritsma_2010} so that the $1$-forms, the edge functions, are specified by our choice of nodes for interpolating the $0$-forms. The straight $0$-forms are interpolated using the the Gauss-Lobatto grid, while the twisted forms use the extended Gauss grid \cite{kreeft2011mimetic}. Because the two grids must be staggered with respect to each other, the grid for the twisted forms contains one more point than that of the straight forms. Hence, the twisted forms are interpolated one polynomial degree higher than the straight forms. The physical domain is divided into elements which are then mapped to the logical domain via a map $F: \hat{\Omega} \to \Omega$. 

The construction of the finite element $L^2$ mass matrices is standard, and the details of its construction are not included. However, the Poincar{\'e} mass matrices are a new concept and we will therefore briefly highlight their construction in $1$D. Suppose we have a non-overlapping partition of the domain: $\Omega = \bigcup_k \Omega_k$. Let $F_k: \hat{\Omega} \to \Omega_k$ be a bijection mapping the reference element to subsets of the physical domain, and let $J_k(\hat{x}) = dF_k/d\hat{x}$ be the Jacobian of this map. Poincar{\'e} duality between $0$-forms and twisted $1$-forms is defined via
\begin{equation}
        \langle u^0, \tilde{v}^1 \rangle = \int_\Omega u^0 \wedge \tilde{v}^1 = \sum_{k=0}^{K-1} \int_{\Omega_k} u^0 \wedge \tilde{v}^1 = \sum_{k=0}^{K-1} \int_{\hat{\Omega}} F_k^* ( u^0 \wedge \tilde{v}^1 ) = \sum_{k=0}^{K-1} \int_{\hat{\Omega}} F_k^* u^0 \wedge F_k^* \tilde{v}^1.
\end{equation}
A similar definition exists for pairing $1$-forms with twisted $0$-forms and so on. Hence, the problem reduces to finding the mass matrix for the reference element. Let $\hat{u}^0_h \in \hat{V}^0_h$ and $\hat{\tilde{v}}^1_h \in \hat{\tilde{V}}^1_h$ where $\hat{V}^0_h$ and $\hat{\tilde{V}}^1_h$ are finite element spaces on the reference element. Then for $\hat{x} \in \hat{\Omega}$, we have
\begin{equation}
    \hat{u}^0_h(\hat{x}) = \sum_i \hat{u}_i^0 \hat{\Lambda}_i^0(\hat{x}) \quad \text{and} \quad \hat{\tilde{v}}_h^1(\hat{x}) = \sum_i \hat{\tilde{v}}_i^1 \hat{\tilde{\Lambda}}_i^1(\hat{x}) \mathsf{d} \hat{x}.
\end{equation}
Hence, the mass matrix for this duality pairing is given by
\begin{equation}
    \mathbb{M}^{01}_{ij} = \int_{\hat{\Omega}} \hat{\Lambda}_i^0(\hat{x}) \hat{\tilde{\Lambda}}_j^1(\hat{x}) \mathsf{d} \hat{x}.
\end{equation}
Similarly, for the other possible Poincaré duality pairings in 1D, we have
\begin{equation}
        \mathbb{M}^{10}_{ij} = \int_{\hat{\Omega}} \hat{\Lambda}_i^1(\hat{x}) \hat{\tilde{\Lambda}}_j^0(\hat{x}) \mathsf{d} \hat{x}, \\
        \quad 
        \tilde{\mathbb{M}}^{10}_{ij} = \int_{\hat{\Omega}} \hat{\tilde{\Lambda}}_i^1(\hat{x}) \hat{\Lambda}_j^0(\hat{x}) \mathsf{d} \hat{x}, \\
        \quad \text{and} \quad
        \tilde{\mathbb{M}}^{01}_{ij} = \int_{\hat{\Omega}} \hat{\tilde{\Lambda}}_i^0(\hat{x}) \hat{\Lambda}_j^1(\hat{x}) \mathsf{d} \hat{x}.
\end{equation}
Evidently, $\mathbb{M}^{10}_{ij} = \tilde{\mathbb{M}}^{01}_{ji}$ and $\mathbb{M}^{01}_{ij} = \tilde{\mathbb{M}}^{10}_{ji}$. While the $L^2$ mass matrices for mapped curvilinear elements involve the metric tensor, because of the presence of the Hodge star operator in the $L^2$ inner product, the Poincar{\'e} mass matrices are metric free objects. 

\begin{figure}
        	\centering
        	\includegraphics[width = 0.8\textwidth]{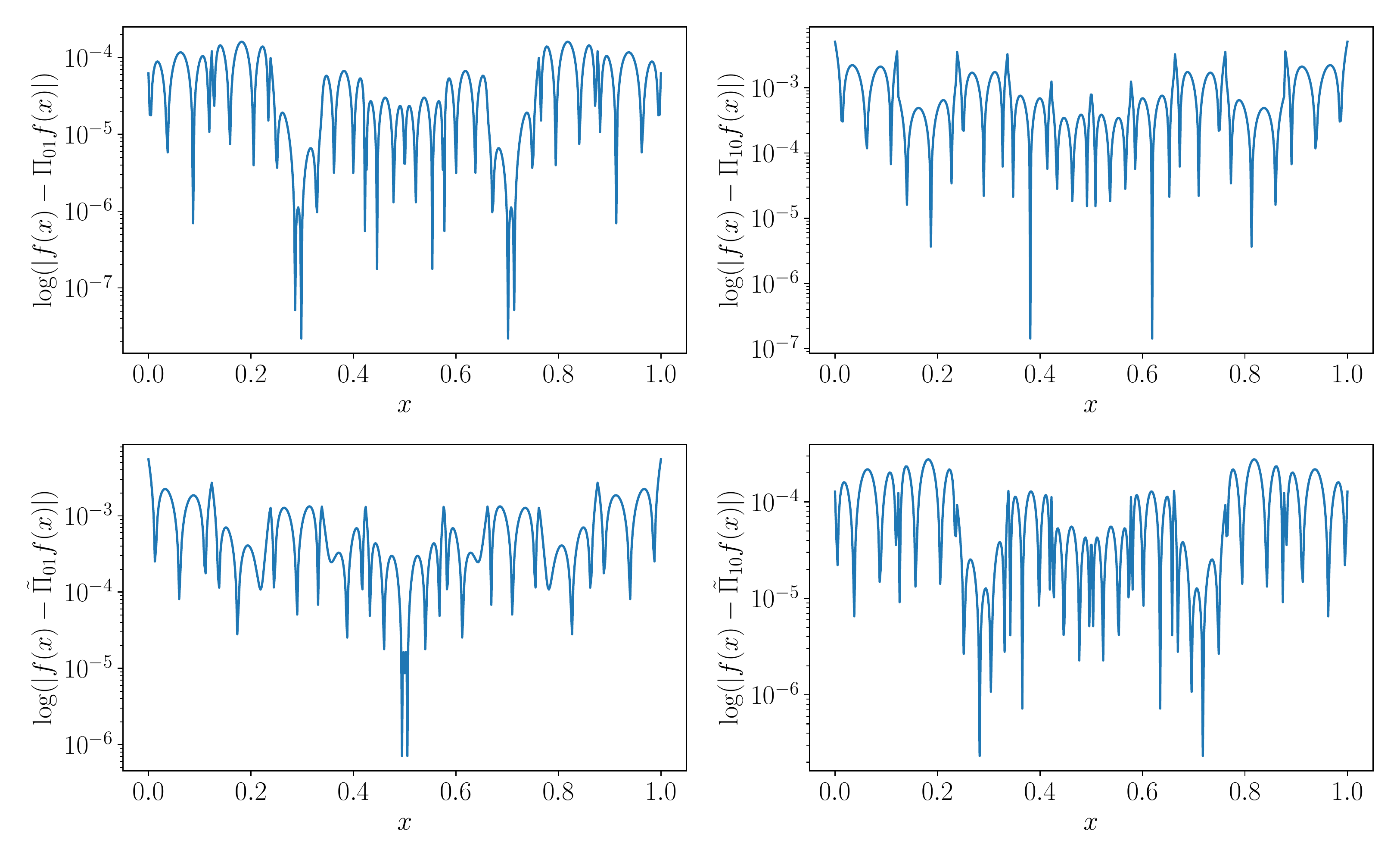}
	\caption{Error in incurred by the dual projections. Here, $\Omega = [0,1]$, the straight $0$-forms are interpolated by $3^{\text{rd}}$ degree polynomials, and $f(x) = \sin(2 \pi x)$.}
	\label{fig:dual_proj}
\end{figure}

To investigate the error incurred by performing the Galerkin projection Hodge star operator, we define the projection-like operators
\begin{equation}
	\Pi_{k, n-k} = \mathcal{I}_k \circ \hollowstar_{k, n-k} \circ \tilde{\bm{\sigma}}_{n-k} \quad \text{and} \quad \tilde{\Pi}_{k, n-k} = \tilde{\mathcal{I}}_k \circ \tilde{\hollowstar}_{k, n-k} \circ \bm{\sigma}_{n-k}.
\end{equation}
These are not true projections as the Galerkin projection Hodge star operator is not invertible. The error incurred by these projections is given in figure \ref{fig:dual_proj}. One can see that the accuracy in each case is limited by the lowest degree polynomial space involved in the projection. 

\subsection*{Convergence study}
\begin{figure}
        	\centering
        	\includegraphics[width = \textwidth]{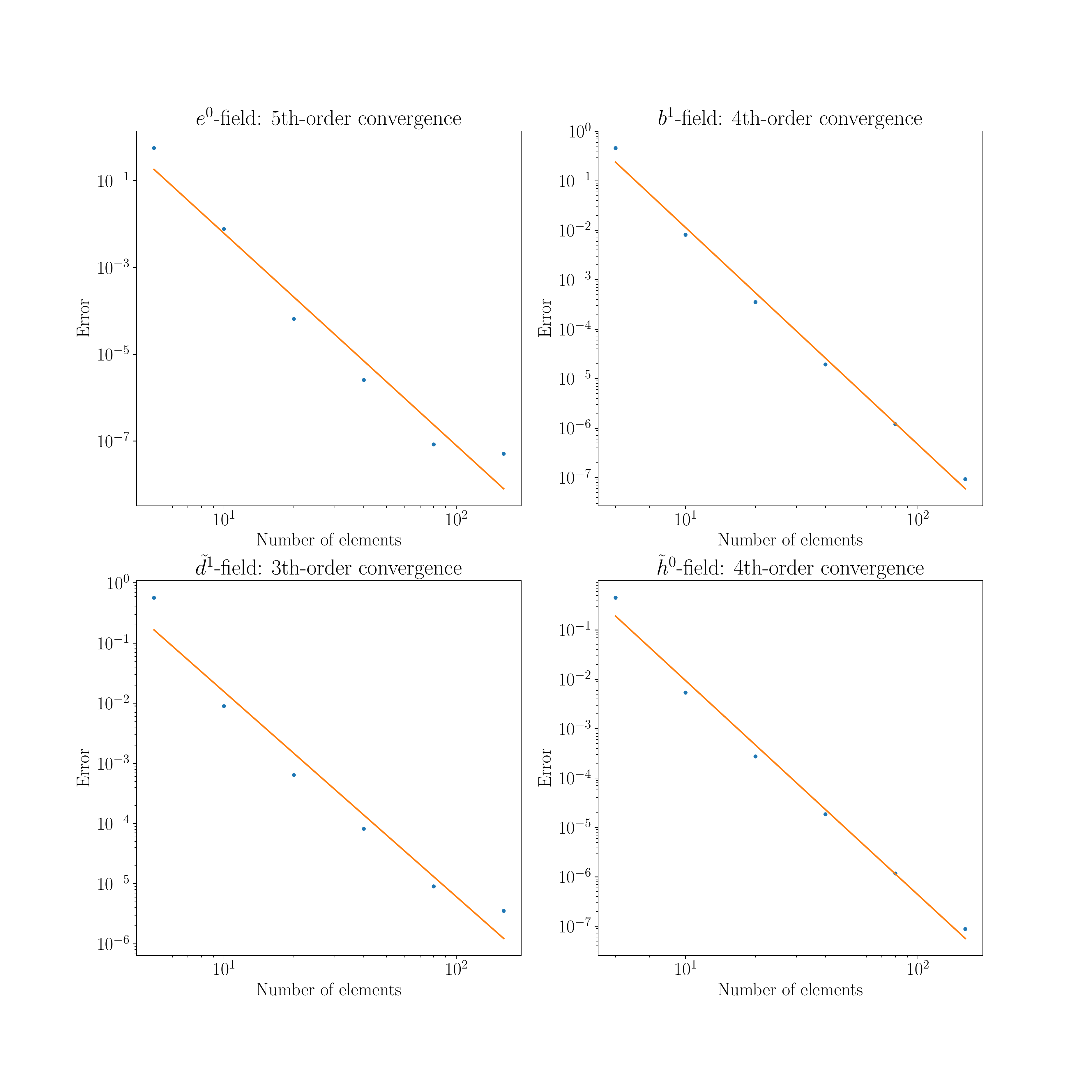}
	\caption{$L^2$ error at time $t=1$ between computed solution and manufactured solution using a $6^{\text{th}}$ order Hamiltonian splitting method \cite{geometric_numerical_integration}, $4^{\text{th}}$ degree polynomials for the straight $0$-forms, $5^{\text{th}}$ degree for the twisted $0$-forms, and one degree lower for the $1$-forms. The number of elements used is $K = 5, 10, 20, 40, 80$, and $160$.}
	\label{fig:conv_rates}
\end{figure}

To begin, we perform a convergence study on the vacuum Maxwell equations with $c=1$ using the method of manufactured solutions. The physical domain is $\Omega = [0,1]$. The manufactured solution is a Gaussian waveform:
\begin{equation}
	E(x,t) = B(x,t) = \exp \left( - \left( \frac{(x - t) \text{ mod } 1 - 1/2}{W} \right)^2 \right)
\end{equation}
where we let $W = 0.1$. The errors between the manufactured and computed solutions at $t=1$ reported in figure \ref{fig:conv_rates}. One can see that, other than an outlier in the error of the $e^0$- and $\tilde{d}^1$-fields at $K = 160$, the convergence rates follow a clear trend. The straight forms achieve the expected convergence rate, but the twisted forms fall short. This is because the method involves projections from higher order polynomial spaces to lower order spaces and the accuracy is limited by the lower fidelity representation. 

\subsection*{Nonuniform dielectric and energy conservation}
\begin{figure}
        	\centering
	\begin{subfigure}{0.85\textwidth}
		\centering
        		\includegraphics[width = \textwidth]{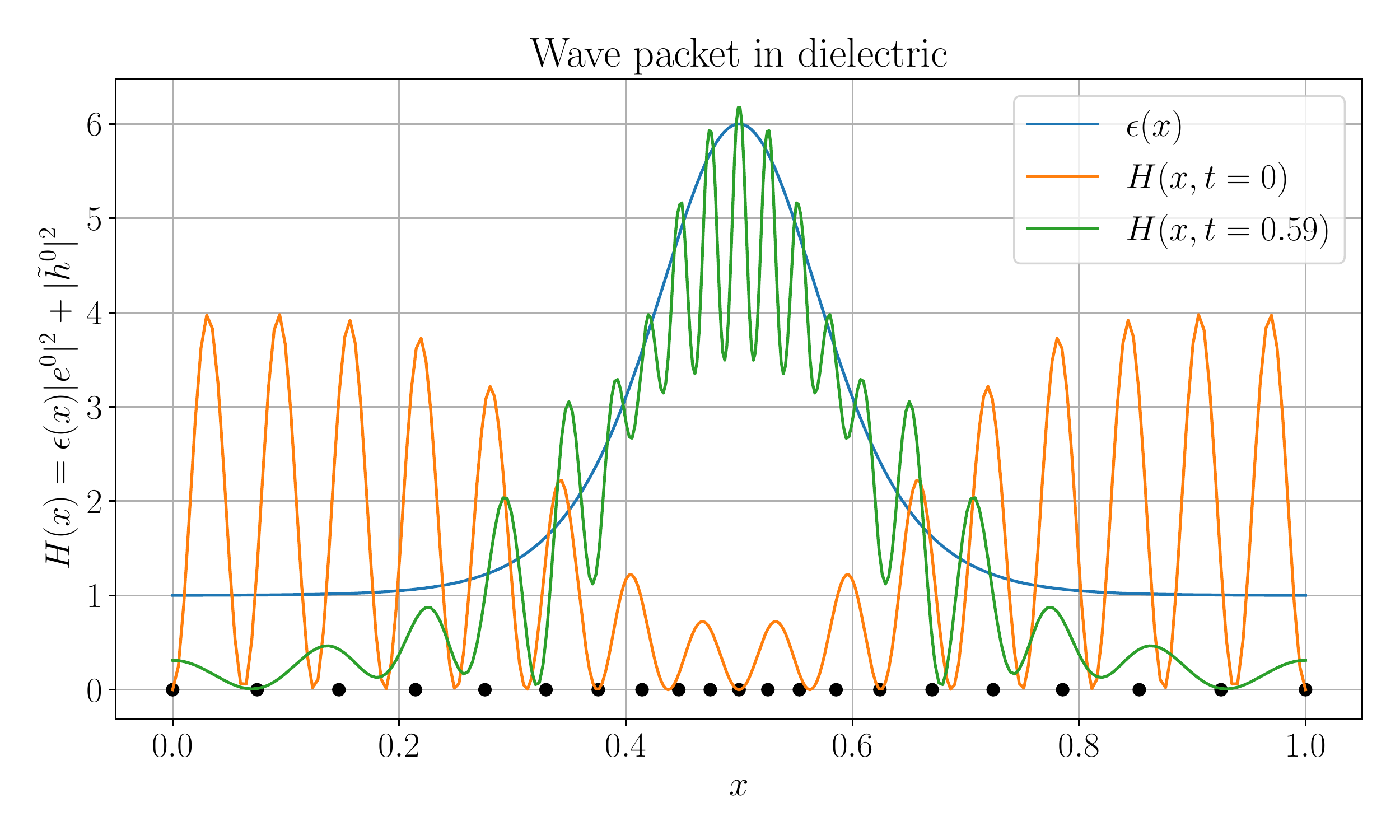}
		\subcaption{Snapshot of the pointwise field energy at $t=0$, at which time the energy is gathered towards the edge of the domain, and $t=0.59$, at which time most of the energy is contained in the dielectric. The black dots indicate the boundaries of each finite element.}
	\end{subfigure}
	
	\begin{subfigure}{0.85\textwidth}
		\centering
        		\includegraphics[width = \textwidth]{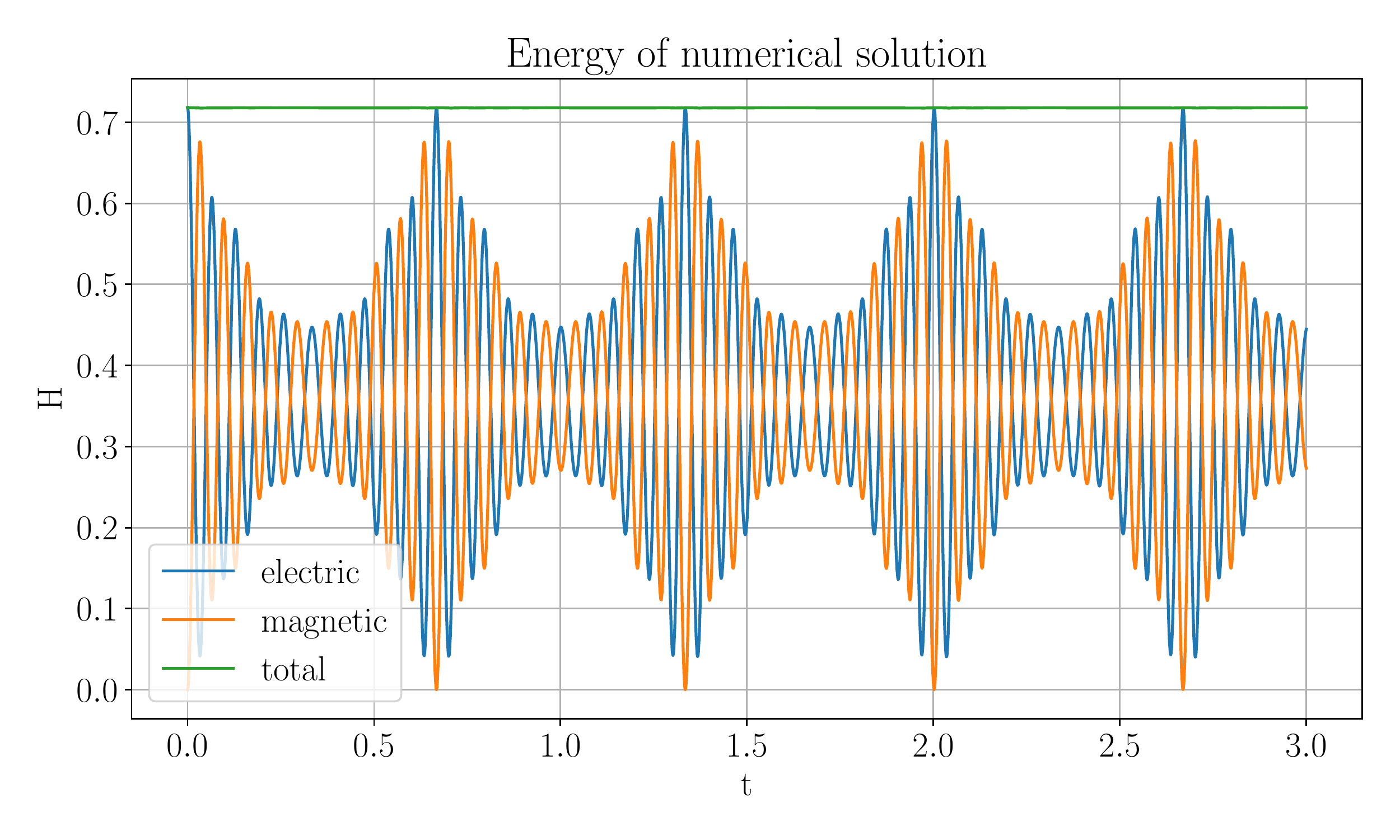}
		\subcaption{The total energy, electric energy, and magnetic energy as a function of time. One can see that the total energy is conserved as desired.}
	\end{subfigure}
	\caption{Solutions with a spatially varying dielectric function. The solution used $K= 20$ elements, $5^{\text{th}}$ order polynomials for the straight $0$-forms, and the $2^{\text{nd}}$ order Strang splitting method \cite{StrangGilbert1968OtCa}.}
	\label{fig:nonunif_dielectric}
\end{figure}

We now consider a test with no magnetization but a spatially varying dielectric function
\begin{equation}
	\epsilon(x) = 1 + 5 \text{sech}^2(10(x -1/2)).
\end{equation}
Hence, the speed of light is slower towards the center of the domain. We should expect more extreme gradients in the center of the domain as the waves tend to be trapped there. Hence, we use the mapping $F(\hat{x}) = (\hat{x} + \sin(\hat{x})/2)/(2 \pi)$ for the grid in order to increase resolution in the center of the domain. See figure \ref{fig:nonunif_dielectric} for the results. The solution behaves qualitatively as it should, the wave travels slower in the dielectric and bunches up there, and the system conserves energy. 

\section{Conclusion} \label{section:conclusion}
The macroscopic Maxwell equations have long been stated in terms of differential forms with the constitutive relations represented by a Hodge star operator \cite{hiptmair_maxwell_cts_disc}. In such models, the permittivity and permeability are typically modeled as linear operators. Likewise, the Hamiltonian structure of the macroscopic Maxwell equations, as a component of a general family of plasma kinetic models, has been known since \cite{morrison:GaugeFreeLifting}. In this model, nonlinear dependence of polarization and magnetization is allowed. In \cite{geoMacroMaxwell}, these two modeling considerations were combined to yield a Hamiltonian formulation of the macroscopic Maxwell equations stated in terms of differential forms. The goal of this paper is to provide a discretization of this model that respects both the geometric structure (namely the structure of the double de Rham complex), and the Hamiltonian structure (variational derivatives, the Poisson bracket, and Casimir invariants). 

Because the geometric structure of the macroscopic Maxwell equations is best stated in terms of the double de Rham complex, discrete structures that accurately capture Poincar{\'e} duality were needed. To this end, the idea of splitting the topological and metric dependence from the split exterior calculus framework of \cite{bauer:hal-02020379} was incorporated into the mimetic spectral element framework of \cite{kreeft2011mimetic}. This yielded a novel Galerkin projection Hodge star operator for the mimetic spectral element method which provides a projection between the straight and twisted de Rham complexes and which decomposes into the product of metric dependent and metric independent matrices. 

Applying this split exterior calculus mimetic spectral element method to the macroscopic Maxwell equations yielded a Hamiltonian system of ODEs in which both Faraday and Ampere's laws were expressed in strong form with the constitutive relations being weakly imposed by Galerkin projection and Gauss's laws are conserved exactly. The numerical results for the 1D Maxwell equations with linear media confirm that the method works and yields the expected behavior. However, it is still necessary to test the method with nonlinear constitutive relations and in higher dimensions. Moreover, future work is needed to use this method on domains that are not periodic.  

\section{Acknowledgements}
We gratefully acknowledge the support of U.S. Dept. of Energy Contract \# DE-FG05-80ET-53088, NSF Graduate Research Fellowship \# DGE-1610403, and the Humboldt Foundation. 


\bibliographystyle{plain} 
\bibliography{dual_mimetic_maxwell}

\end{document}